\documentclass[twocolumn,showpacs,preprintnumbers,amsmath,amssymb]{revtex4}

\usepackage{graphicx}
\usepackage{dcolumn}
\usepackage{bm}

\begin{document}

\title{Height preference and strain in Ag islands on Si(111)$-$(7$\times$7) 
surfaces }

\author{D. K. Goswami$^1$, K. Bhattacharjee$^1$, B. Satpati$^1$, S.~Roy$^1$, G.
Kuri$^2$\footnote{Present address: Paul Scherrer Institute, CH-5232 Villigen
PSI, Switzerland.}, P. V. Satyam$^1$ and B. N.  Dev$^1$\footnote{Electronic
address: bhupen@iopb.res.in}}
\affiliation{$^1$Institute of Physics, Sachivalaya Marg, Bhubaneswar -
751005, India \\
$^2$Hamburger Synchrotronstrahlungslabor HASYLAB at DESY,
Notkestrasse 85, D-22607, Hamburg, Germany}
\date{\today}
\begin{abstract}
    Growth and strain behavior of thin Ag films on Si substrate have been
investigated by scanning tunneling microscopy, cross-sectional transmission
electron microscopy and high resolution x-ray diffraction studies. Ag islands formed on Si at
room temperature growth show strongly preferred heights and flat top. At low
coverage, islands containing two atomic layers of Ag are overwhelmingly formed.
At higher coverages island height distribution shows strong peaks at relative
heights corresponding to an even number of Ag atomic layers. This appears to be
a quantum size effect. Hexagonal disc-like islands with flat top are formed
upon annealing. The annealed film shows two closely-spaced Ag(111) diffraction
peaks $-$ one weak and broad and the other narrow and more intense. The intense
peak corresponds to a shorter Ag(111) planar spacing compared to the bulk
value. This can be explained in terms of changes in the Ag lattice during the
heating-cooling cycle due to thermal expansion coefficient mismatch between Ag
and Si.  
\end{abstract}

\maketitle 

\section{Introduction}

Ag on Si is a nonreactive metal-semiconductor system. The growth of Ag on
Si has been widely studied as a model system. Growth morphology has been
found to depend on the deposition rate and growth temperature.  For Ag
deposition on Si(111)-(7$\times$7) at low temperature (LT), reflection
high energy electron diffraction (RHEED) oscillations were observed up to
many monolayers (MLs) indicating a quasi-layer-by-layer growth
\cite{zhang,roos}. Scanning tunneling microscopy (STM) studies on such
samples also indicated quasi-layer-by-layer growth up to 3.6 ML,
attributed to a lower average island size and higher island density
compared to room temperature (RT) growth \cite{meyer}. At RT deposition it
is generally accepted that the growth follows the Stranski--Krastanov (SK)
or layer-plus-island growth mode
\cite{venables,hanbucken,loenen,chambers}. However, for RT deposition
quasi-layer-by-layer growth has also been observed but only for high
deposition rates ($\sim$30 ML/min) \cite{zhang}. An LT growth followed by
RT annealing leads to yet another growth mode in which 3D plateaulike Ag
islands with a strongly preferred height are observed
on a wetting layer \cite{gavioli}. The
islands increase their number density and lateral extension with coverage
with no change in height, eventually forming a percolated network of the
same preferred height. This behavior was observed where 1 ML to a maximum
of 2.2 ML deposition was studied \cite{gavioli}. This growth mode is 
different from the conventional SK growth mode. The role of electronic
driving force has been suggested to be responsible for this mode of
growth. Zhang and coworkers proposed the {\it electronic growth}
mechanism, in which uniform layer can be grown only over a thickness
window, below or above which the metal film would be nonuniform
\cite{zhang1,suo}. According to their work, very thin films are
destabilized by charge transfer at the interface, films of intermediate
thickness are stabilized by quantum confinement and thick films are
destabilized by stress. Huang et al studied Ag growth on
Si(111)-(7$\times$7) at 50 K followed by annealing at 300 K \cite{22}. By
STM measurements they observed that flat pin-hole free films can be grown
where the amount of material deposited exceeds 6 ML but multilayer pits
are observed for thinner films. Growth at 300 K was found to produce three
dimensional structures. These authors did not explore much higher film
thicknesses to find out if there exists a thickness window above which the
film again becomes nonuniform. To explore this aspect films of higher
thicknesses are to be studied. The authors of ref. \cite{22} studied only
a maximum thickness of 6.4 ML. Moreover, at RT growth whether 
plateaulike islands with a strongly preferred height are formed, as
observed by Gavoli et al. \cite{gavioli} for LT deposition and RT
annealing, has not been explored. In their study the thinnest Ag film on
Si(111)-(7$\times$7) grown at 300 K was 5 ML. Thinner films are to be studied
to reveal any plateaulike islands with a height preference.

	Here we report on our studies of growth of Ag films over a wide
range of thickness on Si(111)-($7\times7$) surfaces for RT growth as
well as RT growth plus annealing at higher temperatures by STM,
transmission electron microscopy (TEM) and high resolution x-ray diffraction (HRXRD)
experiments. Encouraged by the fact that for other metal/Si systems LT
\cite{su} as well as RT \cite{okamota} deposition shows growth of islands
with preferred heights as a consequence of quantum size effect (QSE), we
explore whether a height preference exists for the growth of Ag on Si over
a range of thicknesses for RT growth. Indeed we observe growth of
plateaulike Ag islands with an N-layer (N even) height preference.  These
aspects alongwith the detailed morphology and strain in annealed films are
presented here.
\begin{figure}[h]
\includegraphics[height=13cm]{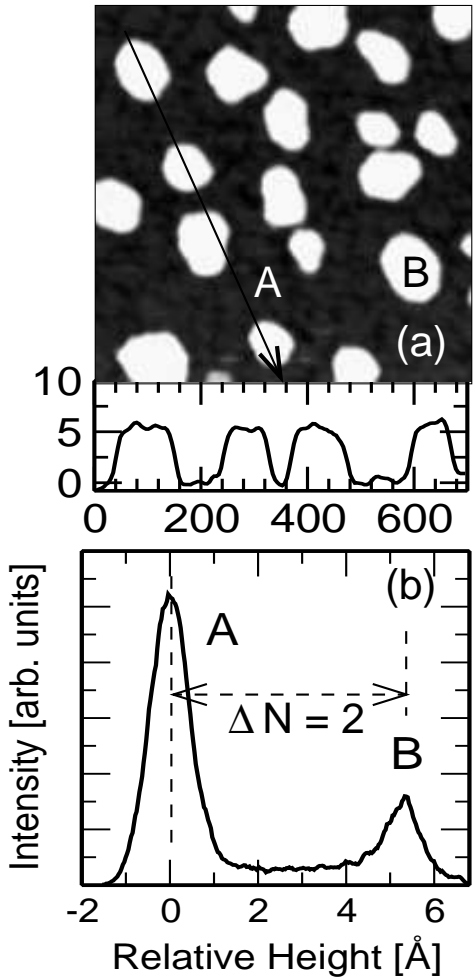}
\caption{
(a) STM image (700$\times$700 $\AA^{2}$)  of a 1 ML Ag film
deposited on a Si(111)-(7$\times$7) surface at RT. Sample bias voltage V$_s$ =
2.1 V, tunneling current I = 0.2 nA. Height profile along the line is shown
(scales are in $\AA$) (b) Island height distribution obtained from the image in
(a), showing the strongly preferred height (peak B)
 corresponding to two atomic layers of Ag(111). Peak A represents the 
Ag wetting layer.
}
\end{figure}

\section{Experimental Details}

Growth and STM measurements were performed in a custom made molecular beam
epitaxy (MBE) chamber coupled with an ultrahigh vacuum (UHV) variable
temperature scanning tunneling microscope (VTSTM, Omicron). This system
has been described elsewhere \cite{dipak}. Base pressure in the growth
chamber was 1$\times 10^{-10}$ mbar. Si(111) samples were cut form n-type
silicon wafers with resistivity in the range 10$-$20 $\Omega$cm. After
introducing into the MBE chamber, substrates were degassed at 600$^\circ$C
for about 12 hours. The sample was then flashed briefly at 1150$^\circ$C
to remove the native oxide and cooled slowly to RT. The (7$\times 7$)
reconstruction was observed on Si(111) surface by STM. Ag was evaporated
from a Knudsen cell (PBN crucible) and deposited onto the
Si(111)-(7$\times 7$) surfaces, kept at RT, at the rate of 2 ML/min. (Some
authors have defined a monolayer of Ag to be equivalent to the normal
surface atomic density of Ag(111), which is 1.5$\times 10^{15}$
atoms/cm$^2$. Others have defined a monolayer to be the equivalent of the
atomic density on an ideal Si(111) surface, which is 0.78 $\times 10^{15}$
atoms/cm$^2$. Here we use the former definition). For the annealed
samples, annealing was performed following deposition in the MBE chamber.
During deposition the chamber pressure rose to 8.5$\times 10^{-10}$ mbar.
The sample was then transfered into the VTSTM chamber for microscopy
measurements at RT. Following STM measurements samples were taken out of
the UHV chamber for HRXRD and TEM measurements.  Film thickness was measured
during growth by a quartz microbalance as well as post-growth Rutherford
backscattering spectrometry (RBS) experiments.  HRXRD measurements with a
rotating anode source and TEM measurements with a JEOL-2010 microscope using
200 keV electrons were made in Institute of Physics. HRXRD
measurements were also performed with synchrotron radiation at HASYLAB at
DESY in Germany at the bending magnet beamline ROEMO-I.

For HRXRD measurements with a rotating anode source, we have used Mo
K$_{\alpha 1}$ X-rays ($\lambda = 0.709 \AA$) monochromatized by a
symmetrically cut Si(111) crystal. For experiments with synchrotron
radiation (SR) a pair of symmetrically cut Ge(111) crystals were used in a
double-crystal monochromatator. A beam entrance slit (S1: vertical opening
$ 70 \mu$m, horizontal opening $ 0.5$ mm) 46 cm before the sample and
another slit(S3: vertical opening 200 $\mu$m) before the detector at a
distance of 43 cm from the sample were used. An antiscattering slit (S2:
vertical opening 80 $\mu$m) was used between the sample and the entrance
slit. During sample alignment smaller vertical opening of the slits were
used (S1: 30 $\mu$m, S3: 100 $\mu$m).  X-rays of a longer wavelength
($\lambda = 1.033 \AA$) were used for the SR experiments.
\begin{figure}[h]
\includegraphics[height=8cm]{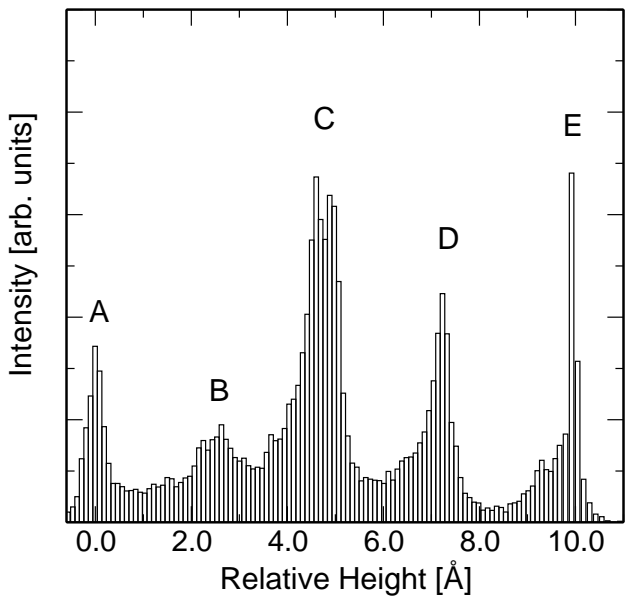}
\caption{
 Height distribution obtained from of a STM image of a 2 ML Ag
film. Peak A corresponds to the wetting layer and B, C, D and E correspond to
islands of monolayer, bilayer, trilayer and quadrilayer heights, respectively.
Islands of double-layer and four-layer heights are dominant.
}
\end{figure}

\section{Results and discussions}

\subsection{Growth and morphology}

For a thin (1 ML) Ag layer grown on a Si(111)-$(7\times7)$ surface at RT,
a STM image is shown in Fig.1(a). A height scan along the line marked in
Fig.1(a) is shown below the image. Growth of flat-top islands with a
preferential height is evident from the height scan. Indeed all the
islands seen in Fig.1(a) have the same height. Height distribution of Ag
islands obtained from the micrograph in Fig.1(a) is shown in Fig.1(b). The
dark background area (marked `A') in Fig.1(a) corresponding to peak A in
Fig.1(b) is the first Ag wetting layer on Si(111). The bright features
(marked `B') in Fig.1(a) corresponding to peak B in Fig.1(b) are Ag
islands with flat top and a height of 5.4 $\AA$ from the Ag wetting layer.  
Assuming that the spacing between atomic layers in the Ag islands is close
to the corresponding value in bulk Ag along the [111] direction, each
island contains two atomic layers of Ag. The wetting layer contains about
0.5 ML Ag \cite{gavioli}, the remaining deposited Ag grows as islands on
the wetting layer.  So the double-layer height preference of the Ag
islands is obvious. This height preference of Ag islands for RT growth of
Ag on Si(111)-($7\times7$) surfaces is quite robust.  Among several STM
images, in one image similar to Fig.1(a), among islands of double-layer
height only one island of a height corresponding to three atomic layers of
Ag was found. We observed no Ag islands of one atomic layer height.  The
root-mean-square roughness on any given island is $\sim$0.2 $\AA$ while
that obtained from the whole image [Fig.1(a)] is 2.2 $\AA$. The
double-layer height (5.4 $\AA$) preference of Ag islands for the
Ag/Si(111)-$(7\times7)$ system was earlier observed only for LT (150 K)
deposition followed by RT (300 K) annealing for deposition between 1 and
2.2 ML Ag \cite{gavioli}.  

This novel growth mode was qualitatively
explained in terms of the {\it electronic growth} mechanism
\cite{zhang1,suo}, wherein the quantized electrons in a layer can
influence the morphology. As we observe here, for the Ag/Si(111) system LT
growth followed by RT annealing is not a necessary condition for the
formation of islands of preferred heights. They are also formed in RT
deposition. Moreover, height preferences with larger heights have been
observed in other metal/silicon systems \cite{su,okamota}. We investigated
growth of Ag layers of various thicknesses (1$-$60 ML). For 2 ML Ag films
preference for growth of islands of N-layer (N$\ge 2$) height is observed.
A height distribution plot obtained from a STM image of a 2 ML film is shown in Fig.2.
There is a strong tendency of height preference in units of bilayer height
[N = 2 (A--C), 4 (A--E)].  From Fig.2 we also notice the existence of
monolayer B and trilayer D heights; however, their intensities are
smaller. The monolayer height peak B arises from the height difference
between C and D, i.e., growth of monolayer height islands on the already
formed double-layer height islands. Direct growth of monolayer height
islands on the wetting layer A is very rarely observed. For the growth of 4
ML onwards, the Ag layer forms a percolating structure. 
\begin{figure}[h]
\includegraphics[height=7cm]{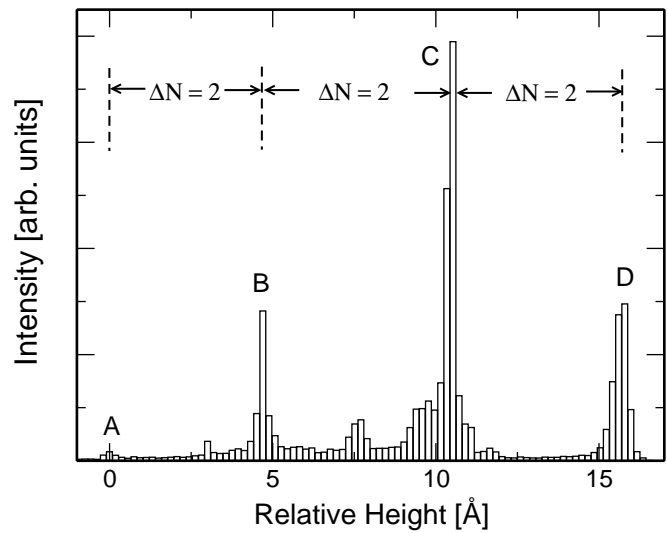}
\caption{
 Height distribution obtained from a STM image (350 $\times 850
\AA^2$) of a 5 ML Ag film. N-layer heights of islands (B: N = 2, C: N = 4, D: N
= 6) for even values of N are strongly preferred.
}
\end{figure}
The height
distribution from a 5 ML Ag film is shown in Fig.3. Growth of islands of
N-layer (N even: 2, 4, 6) height is prominent.  The peaks A, B, C, D in
Fig.3 correspond to the wetting layer, N = 2, N = 4, N = 6 islands. These
heights are seen in line scans through the STM image (not shown).  STM
images from 20 ML, 30 ML and 40 ML Ag films are shown in Fig.4. The
roughness obtained from the STM image in Fig.4(c) is 6.2 $\AA$, while that
only on the flat features is $\sim$ 0.2 $\AA$. Small hexagonal islands
appear to grow on the outer layers beginning around 30 ML thick films.
These hexagonal islands become more prominent for 40 ML and 60 ML (shown
later) films. Height distribution in the 40 ML film is shown in Fig.5.
Height preference for even-N is observed. Peak A in Fig.5 is not from the
wetting layer. The wetting layer has been covered by further growth on it.
If we assume the height corresponding to A to be N$_0$-layer then we
observe strong preference for N$_{0}+$2, N$_{0}+$4 and N$_{0}+$6 layers in
Fig.5. These heights can be appreciated from the height scan and the grey
scale in the marked area in Fig.4(c). Apparently the N-layer height
preference with an even value of N is an effect of quantum confinement of
electrons in the Ag islands. To our knowledge in the existing literature
neither theoretical nor experimental results (except for N = 2
in LT growth followed by RT annealing \cite{gavioli}) on the height 
preference in Ag(111) films on Si(111)
\begin{figure}[h]
\includegraphics[height=18cm]{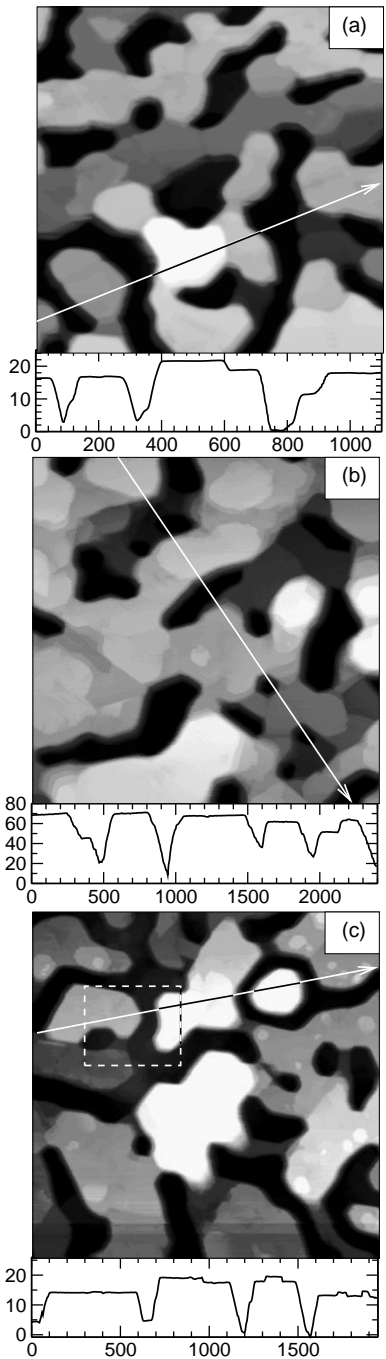}
\caption{
 STM images of (a) 20 ML (1000$\times$1000 $\AA^{2}$),
(b) 30 ML (2000$\times$2000 $\AA^{2}$) and (c) 40 ML (2000$\times$2000
$\AA^{2}$) Ag films. Height profiles along the marked lines are shown (scales
are in $\AA$). Small hexagonal islands on the top layer begins to
grow for $\gtrsim$ 30 ML films. These islands are more abundant on 40 ML
films.
}
\end{figure}
\begin{figure}[h]
\includegraphics[height=7cm]{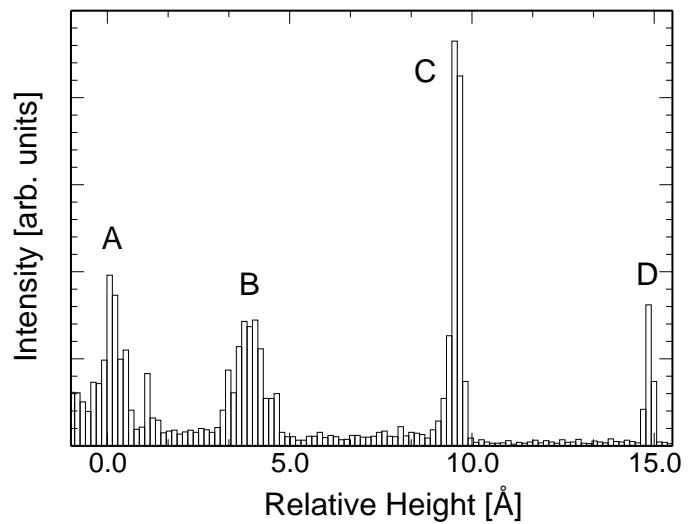}
\caption{
 Height distribution obtained from the marked area with dashed
lines in Fig. 4(c). Relative heights apparently correspond to A--B: $\Delta$N =
2, A--C: $\Delta$N = 4,
A--D: $\Delta$N = 6. Here the peak A does not
corespond to the wetting layer. Relative height for a given $\Delta$N
may somewhat depend on the value of N ( see text for details).
}
\end{figure}
substrates are available. Recently for Pb islands grown on Si(111)  
substrates, Okamoto et al. observed a preference for islands with height
differences ($\Delta$N) equivalent to an even number of Pb atomic layers
\cite{okamota}. Surface energy has been found to be lower for a height
containing an even number of atomic layers, compared to the neighboring
heights containing an odd number of layers. For Ag films on Fe(100)
surfaces, in general an N-layer film, though stable at low temperature,
was found to bifurcate into a film with N$\pm$1, i.e. $\Delta$N=2, around
400 K \cite{luh}. Some theoretical attempts have been made to understand
the N--dependent stability of metal films using the quantization condition
(the phase accumulation model):

\begin{equation}
2k_z(E)Nd_0 + \phi_1(E) + \phi_2(E) = 2n\pi
\end{equation}

where $k_z$ is the component of the wave vector prependicular to the film
surface, $k_z(E)$ is determined by the band structure, $d_0$ is the
interlayer spacing and $n$ is an integer.  $\phi_1(E)$ and $\phi_2(E)$ are
the energy-dependent phase shifts of the electronic wave function upon
reflection at the two boundaries of the film \cite{wei}. For the
Ag(111)/Si(111) case this would depend on the Ag band structure (energy
dispersion along [111] direction) and the phase shifts at Ag/vacuum and
Ag/Si interfaces.  Although for a direct comparison, calculation for the
Ag/Si(111) system would be necessary, it is evident that quantum size
effect (QSE) is important in leading to the observed height preference of
Ag islands.

It is interesting to note that the bilayer height on the wetting layer for
the thinnest Ag film (1 ML) is about 5.4 $\AA$ (Fig.1). (The same value of
the double-layer height was observed by Gavioli et al.\cite{gavioli} for
film thicknesses between 1 and 2.2 ML.  These authors did not study
thicker films).  However, for the thicker films (see figures 2, 3 and 5)
we observe some variation in the bilayer separation. [It should be noted
that the actual value depends somewhat on the magnitude of the sample bias
voltage and its sign. However, the trend remains the same]. While for Ag
islands on Si no report is available on thickness relaxation depending on
the total number of atomic layers in an island, for Pb islands on Si(111)
surfaces an oscillatory thickness relaxation was observed depending on the
number of atomic Pb layers in the islands. This oscillatory thickness
relaxation has been shown to be correlated with quantized electronic
states in the Pb islands \cite{su}. The thickness relaxation has been
characterized in terms of the deviation from ideal thickness, defined as
$\Delta t = t_N - Nd_0$, where N is the number of atomic layers on the
wetting layer, $d_0$ is the ideal interlayer spacing and $t_N$ is the
thickness of the island. For small $N$ the deviations were found to be
large [for example, $\Delta t = -0.5\AA$, $0.2\AA$ and $-$0.4 $\AA$ for $N
= 5$, $N = 6$ and $N = 7$ respectively].  The amount of deviations also
apparently depend on whether the islands are isolated individual islands
or islands of several tiers. (For Ag islands we show the characteristic
differences for these cases later).  With increasing thickness the island
height variation tends to diminish.

For as-deposited 60 ML films, growth of hexagonal islands is prominent as
seen in the STM image of Fig.6. A height scan along the line in Fig.6(a)
across multi-tier islands is shown in Fig.6(b). Single-layer height at the
island edges are prefered. It is seen from Fig.6(b) that step heights for
the inner layers are $\sim$ 2.6 $\AA$, while those of the outer layers are
$\sim$ 3 $\AA$ -- both much larger than the ideal Ag(111) layer spacing
(2.36 $\AA$). [ It should be noted that height measurement by STM is
affected by both structural and electronic contribution]. As we will show
later for annealed samples, where both individual and multi-tier islands
are observed, multi-tier island edges prefer single-layer height (like that
in Fig.6) whereas individual islands prefer double-layer or in general
even-layer height. This trend has also been observed for Pb islands on
Si(111), where N values, which are practically absent in isolated islands,
are observed for multi-tier islands showing monolayer steps \cite{su}.
\begin{figure}[h]
\includegraphics[height=13cm]{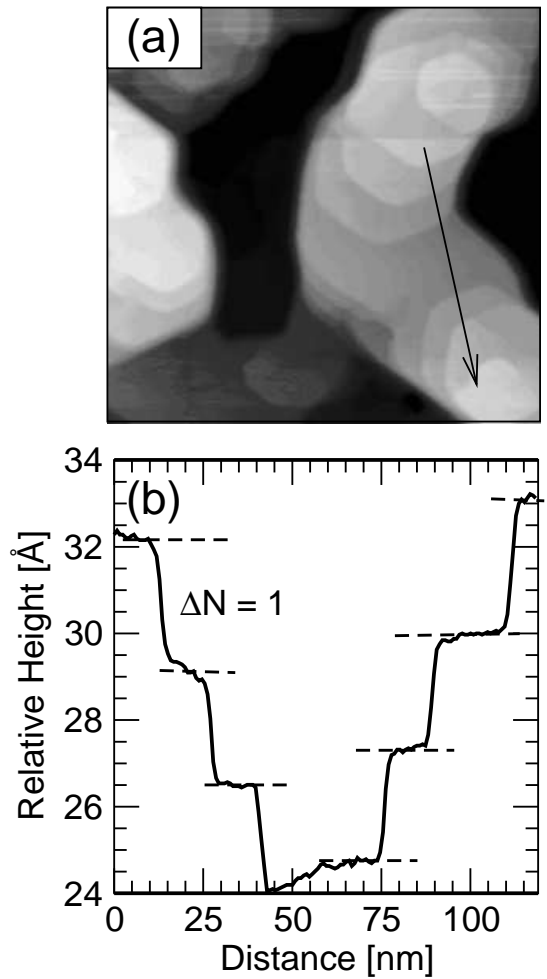}
\caption{
(a) STM image ($3000 \times 3000 \AA^2$) of an as-deposited 60
ML Ag film. Multi-tier islands, present but not prominent on the 40 ML film,
are very prominent here. (b) a height scan along the line in
(a). Monolayer ($\Delta$N = 1) steps are found to be dominant for these
multi-tier islands.
}
\end{figure}

In the standard SK growth in RT deposition for film thicknesses above the
wetting layer, formation of 3D multilayer Ag islands have been observed
\cite{loenen}. Here in RT deposition for thin Ag films (1 ML) we have
observed formation of 3D Ag islands with flat top and a strong preference
for a double-layer height.  Growth of 3D Ag islands with a flat top and a
preferred double-layer height was earlier observed only for 1$-$2.2 ML Ag
films deposited at LT (150 K) followed by annealing at RT (300 K). For RT
deposition this feature of Ag island growth on Si with a height preference
has not been reported earlier. However, for Pb islands on Si(111), island
height preference indicating quantum size effect was observed for growth
at 170$-$250K \cite{su}, 273K \cite{altfeder} and even at RT
\cite{okamota}. For higher film thicknesses (4 ML onwards), we observe the
growth of a percolated Ag layer with a flat top feature on the islands
with a height preference in units of double-layer height.
\begin{figure}[h]
\includegraphics[height=7cm]{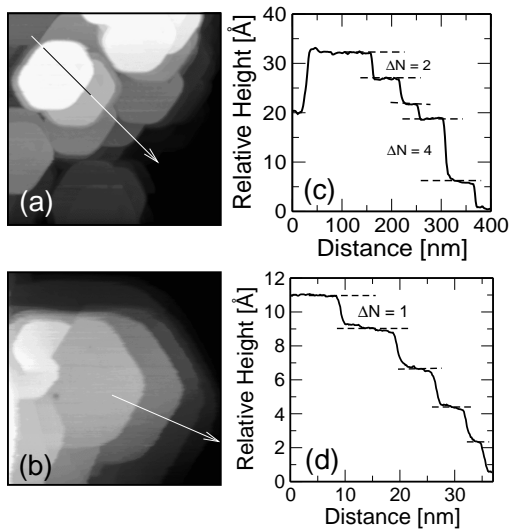}
\caption{
 STM images (a, b) of a 40 ML Ag film annealed at
700$^{0}$C. Prominent hexagonal disc-like islands are seen to have been
formed. There are significant differences in the islands in (a) and
(b). In (a) islands are apparently independent islands while in (b)
smaller islands on progressively larger islands in a multi-tier form
are seen. (c) Height scan along the line in (a) shows strong
preference for height differences corresponding to an even number of
Ag atomic layers. (d) Height scan along the line in (b) shows
monatomic height on the island edges similar to what is seen in Fig.6
for multi-tier islands. Note the large difference in lateral
dimensions in (a) and (b). Individual hexagonal islands are much larger.}
\end{figure}

	The RT-deposited 40 ML film, when annealed at 700$^\circ$C,
nanostructural hexagonal disc-like Ag islands are formed as seen in the
STM micrograph of Fig.7(a). [Very few hexagonal islands appear in the
as-deposited film (compare Fig.4(c))].  According
to kinetic Monte Carlo simulation results, islands can grow on a substrate
of threefold symmetry in various shapes depending on growth temprature $-$
hexagonal shape being one of them \cite{kmc}. We have also observed
triangular island growth in a Ag film on Si(111) annealed at 600$^\circ$C
\cite{dipak}.  A height scan along the line in Fig.7(a) is shown in
Fig.7(c). We notice that the hexagonal disc-like Ag islands have a
preference for a height difference, where $\Delta$N is even. This appears
to be the case when the islands are individual islands.  However, when the
islands are multi-tier islands, i.e.  small islands growing on larger
islands, the edges prefer to have monolayer steps, as seen in Fig.7(b) and
its corresponding height scan in Fig.7(d).  Differences between individual
and multi-tier islands have also been observed for Pb islands on
Si(111)$-(7\times7)$ surfaces \cite{su}, although for much thinner films.
For the Ag films of Si(111), up on annealing, growth of individual islands
with $\Delta$N = 2 or 4 becomes more prominent compared to multi-tier
islands with $\Delta$N = 1. This indicates that $\Delta$N = 2 or 4 is
energetically more favorable. For Pb islands on Si(111) as well, up on
annealing $\Delta$N = 1 tends to disappear in favor of $\Delta$N = 2
islands \cite{su}. For Pb/Si(111) this has been attributed to quantum size
effect. For Ag/Si(111) theoretical results would be necessary for a direct
comparison.

	Within our limited search over selected thicknesses (1, 2, 4, 5,
10, 30, 40 and 60 ML) we have not observed uniform film growth. An
extended search would be necessary to explore if there at all exists a
thickness window over which film of uniform thickness can grow at RT or
any other growth temperature.
\begin{figure}[h]
\includegraphics[height=6.5cm]{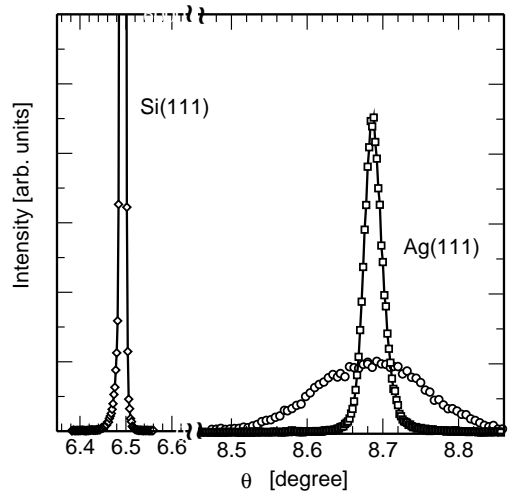}
\caption{
 X-ray diffraction data obtained from 40 ML Ag films with
x-rays of $\lambda$ = 0.709 $\AA$ from a rotating anode source (circle:
as-deposited film, square: 700$^{0}$C-annealed film). Epitaxy of Ag(111) layer
improves up on annealing as the narrowing of the Ag(111) peak indicates.
}
\end{figure}

\subsection{Strain}

High resolution x-ray diffraction (HRXRD) measurements were made 
with a rotating anode X-ray
source (Mo K$_{\alpha 1}$, $\lambda = 0.709 \AA$) on the
Ag layers.  Typical HRXRD results ($\theta$--2$\theta$ scan) from a 40 ML Ag
film are shown in Fig.8. Epitaxial growth of Ag(111) layer is seen from
the Ag(111) diffraction peak. No other Ag peaks were observed. The
diffraction peak from the as-deposited film is broad, which can be
attributed mainly to a mosaic spread. Mosaic spread in Ag(111) layers deposited
on Si(111) at RT has been observed earlier \cite{luo,sundar1}. Mosaic
spread decreases upon annealing.  HRXRD results from a Ag layer annealed at
700$^\circ$C (30 min.) shows a much narrower Ag(111) peak indicating a
reduction of mosaic spread and improvement of epitaxy upon annealing.  
The Ag film, annealed at 700$^\circ$C, as mentioned before, form hexagonal
disc-like islands (see Fig.7). Much thicker Ag islands are formed in this
annealed sample. TEM measurements reveal this feature. Typical
cross-sectional TEM micrographs and a transmission electron diffraction
pattern are shown in Fig.9.  The island in Fig.9(a) is $\sim500\AA$ thick
$-$ more than five times the deposited nominal thickness. The selected area
diffraction pattern taken from the region shown in Fig.9(a) 
shows diffraction from both Si and Ag indicating
Ag(111)~$\parallel$~Si(111) and Ag[110]~$\parallel$~Si[110]. An isolated Ag
island on Si of $\sim$ 2000$\AA$ thickness is seen in Fig.9(c). In
addition to the reduction in mosaic spread in the annealed film, growth of
thick islands also may be partly responsible for narrowing of the Ag(111)
peak in Fig.8.
\begin{figure}[h]
\includegraphics[height=13cm]{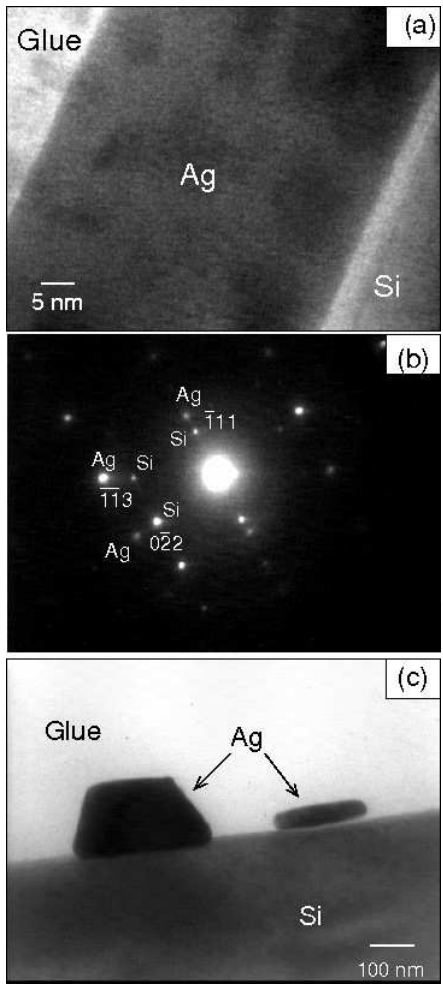}
\caption{
 Cross-sectional TEM micrograph (a) and a transmission electron
diffraction pattern (b) from a 40 ML Ag film annealed at 700$^{0}$C. Epitaxy of
Ag(111) with Si(111) is seen from the diffraction pattern. (c) An island about
20 times as thick as the nominal deposited thickness is seen to have formed.}
\end{figure}
\begin{figure}[h]
\includegraphics[height=6.5cm]{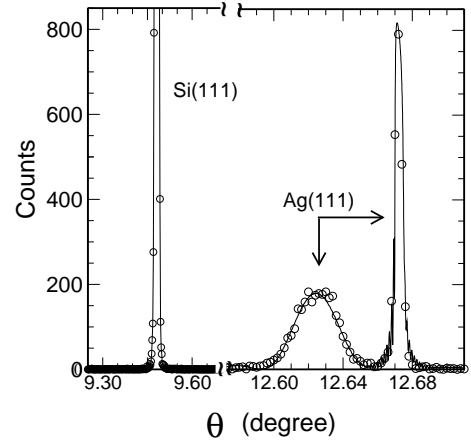}
\caption{
 HRXRD data from a 40 ML Ag film annealed at
700$^{0}$C, obtained with synchrotron radiation ($\lambda$ = 1.033
$\AA$). Two distinct Ag(111) peaks are observed with a small angular
separation $\Delta\theta$ = 0.05$^{0}$.  The $d$-spacing
coresponding to the stronger Ag(111) peak is less than that of bulk Ag 
(see text for explanation). The narrowness of the peak is consistent
with the formation of thick islands seen in Fig.9.
}
\end{figure}

	HRXRD experiments were also performed on a 700$^\circ$C-annealed 40
ML film with monochromatized ($\lambda = 1.033 \AA$) synchrotron
radiation. The result (Fig.10) shows two distinct Ag(111) peaks alongwith
the Si(111) diffraction peak. The results in Fig.10 were obtained with
$\theta$-steps of 0.002$^\circ$. However, we have also verified the
authenticity of the Ag double peak structure by repeating experiments with
0.001$^\circ$ steps.  The sharp peak and the wide peak correspond to
Ag(111) planner spacings of 2.355 $\AA$ and 2.364 $\AA$ ($\Delta d = 0.09
\AA$), respectively ($\Delta\theta = 0.05^\circ$ between the peaks). The
angular positions of the Ag peaks have also been verified with reference
to the Si(111) peak position. The sharpness of the peak at larger
$\theta$, i.e.  for {\it d}=2.355 $\AA$ indicates that this component
comes from a much thicker layer. As we have seen from Fig.9, much thicker
islands, compared to the deposited nominal thickness, are formed in the
annealed sample. The broader weak peak apparently comes for thinner
islands. The Ag peaks have been fitted using the Takagi-Taupin formalism
based on dynamical theory of x-ray diffraction \cite{bartels}. The 
theoretical curve for the narrow Ag(111) peak has interference fringes near
the base, which are barely visible in this scale.

     We now explore the possible origin of the sharp Ag(111) peak
corresponding to a shorter plannar spacing. Lattice constants of Ag and Si
are {\it a}$^{Ag}$=4.085$\AA$, and {\it a}$^{Si}$=5.431$\AA$ ({\it
d}$^{Ag}_{111}$=2.359$\AA$ and {\it d}$^{Si}_{111}$=3.136$\AA$). The
lattice mismatch is $\sim$25\%. However Ag is found to grow epitaxially on
Si \cite{luo,sundar1,legoues}. This happens via coincidence site lattice
matching as 4{\it a}$^{Ag}\approx$3{\it a}$^{Si}$ and the mismatch is just
0.43\% (4{\it d}$^{Ag}_{111}>$3{\it d}$^{Si}_{111}$) \cite{zur,legoues}.
In this case an in-plane compression of the Ag layer would lead to a
perpendicular strain leading to {\it d}$^{Ag}_{111}$(film)$>${\it
d}$^{Ag}_{111}$(bulk).  However, what we observed is {\it
d}$^{Ag}_{111}$(film)$<${\it d}$^{Ag}_{111}$(bluk) as one component, the
other being {\it d}$^{Ag}_{111}$(film)$>${\it d}$^{Ag}_{111}$(bluk). A
possible explanation for the shorter {\it d}-spacing (2.355 $\AA$) in Ag
can be given in terms of coefficients of linear thermal expansion of Ag
(18.9 $\times 10^{6}$/K) and Si (2.6 $\times 10^{6}$/K) \cite{web}. As the
expansion coefficient of Ag is about seven times that of Si, the Ag
lattice being pinned at the Ag/Si interface would expand faster in the
surface-normal direction enhancing strain in the Ag layer. For the
Ag/Si(111) system at 700$^\circ$C the strain in the Ag layer would be much
larger compared to that at RT. If we assume that this larger strain would
be relaxed by introducing dislocations at the interface and a fully
relaxed Ag layer would form at 700$^\circ$C, the fully relaxed Ag layer
would have {\it d}$^{Ag}_{111}=2.359\AA$. As the thermal expansion
coefficient of Ag is much higher than that of Si, in cooling the sample
down to room temperature, the Ag layer would shrink at a higher rate than
Si. As the in-plane Ag lattice would be pinned at the interface to that of
Si, which shrinks at a slower rate, the Ag layer would have a larger
shrinking in the surface-normal direction leading to a Ag(111) planar
spacing in the surface-normal direction smaller than that of bulk Ag. This
is schematically illustrated in Fig.11. A thick Ag layer is likely to
undergo this process. This is also consistent with the narrowness of the
Ag(111) peak corresponding to the shorter {\it d}-spacing ({\it
d}$^{Ag}_{111}=2.355\AA$). The lattice contraction is 0.17\% with respect
to the bulk $d_{111}$ spacing. The weak broader Ag(111) peak is likely to
be due to much thinner islands, which would be able to accommodate the
strain due to thermal expansion mismatch without introducing dislocations.
The presence of a thinner island is also seen in Fig.9(c). There are
actually thin islands with a thickness variation. The broader peak
corresponds to a lattice expansion of 0.21\% compared to the bulk
$d_{111}$ spacing.  For $\lambda = 1.033 \AA$, the angular separation
($\Delta\theta$) between the two Ag(111) peaks, as seen from Fig.10, is
0.05$^\circ$.  For $\lambda = 0.709 \AA$ this separation would be
$\Delta\theta = 0.03^\circ$ and with our laboratory set up it would be 
more difficult to observe these peaks as separated peaks. 

\begin{figure}[h]
\includegraphics[height=4.6cm]{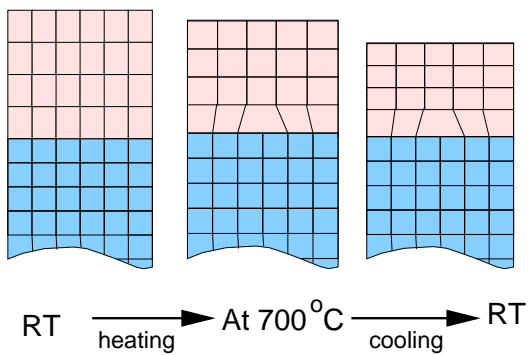}
\caption{
The shortening of the perpendicular $d$-spacing in the
annealed film compared to  the corresponding bulk value is schematically
illustrated. See text for details. 
}
\end{figure}


There are reports that desorption of Ag from a Ag/Si system starts around
550$^\circ$C under UHV conditions \cite{ino}. Earlier we studied growth of
Ag(111) layers ($\sim1000\AA$) on Si(111) and their annealing behavior
under high vacuum (HV) condition by RBS/channeling 
experiments \cite{sundar1}. We have studied Ag/Si
annealing (30 min) under high vacuum upto 800$^\circ$C. At
600$^\circ$C, we observed very little loss of Ag. At 700$^\circ$C there
were about 25\% loss of Ag $-$ mainly from grain boundaries without any
detectable change of actual film thickness. At 800$^\circ$C (30 min) Ag films
were completly desorbed \cite{sundar2}. In the present case we expect that
some desorption of Ag has taken place in the annealed film.

\section{Summary and Conclusions}

	We have studied growth of Ag on Si(111)-$(7\times7)$ surfaces at
room temperature. Initial deposition of $\sim$ 1 ML Ag produces islands
with a strongly preferred height of two atomic layers of Ag on the wetting
layer.  Thicker films show a tendency of growth of N-layer islands, where
N is even. This appears to be a consequence of electron confinement in the
metallic film.  As-deposited thicker films ($>$20 ML) show the formation
of hexagonal islands at the top. These multi-tier islands have monatomic
step edges.  Annealed films show the formation of hexagonal disc-like
islands. Individual islands show double-layer height at the edges, while
multi-tier islands show monatomic height edges as in the as-deposited
films.  While the as-deposited films are epitaxial with a mosaic spread,
the epitaxy improves up on annealing. An annealed 40 ML film has been
found to show two closely-spaced Ag(111) peaks. The strong peak
corresponding to shorter {\it d}-spacing compared to the bulk value can be
explained by the perependicular contraction up on cooling of the Ag lattice,
which is presumably relaxed at high temperature during annealing, as a
consequence of thermal expansion mismatch between Ag and Si.

	Within our limited search we have not observed, for RT growth, any
thickness window for the uniform layer growth $-$ a prediction from the
electronic growth mechanism. Obviously an extended search over film
thickness and growth temperature would be necessary to explore this
aspect.

	The fact that bilayer growth is prefered even at the outer layers
of relatively thick films needs further theoretical attention. Ag growth
on Si(111)-(7$\times$7) at RT cannot be simply classified within the three
commonly known growth modes. The elctronic confinement apparently plays a
role in determining the film morphology. To what thickness electron
confinement effect is operative is also an important aspect to be
investigated.

\section{Acknowledgements}

We acknowledge the help provided by the HASYLAB staff. BND would like to thank
Prof. J. R. Schneider for extending his support.


\begin{thebibliography}{99}
\bibitem{zhang} Z. H. Zhang, S. Hasegawa and S. Ino,  Phys. Rev. 
        {\bf B55}, 9983 (1997)
\bibitem{roos}K. R. Roos and M. C. Tringides, Surf. Sci. {\bf 302}, 
        37 (1994)
\bibitem{meyer}G. Meyer and K. H. Rieder,  Appl. Phys. Lett. {\bf 64}, 
        3560 (1994)
\bibitem{venables} For reviews, see J. A. Venables,  Surf. Sci. 
        {\bf 299/380}, 789 (1994); C. Argile and G. E. Rhead,  Surf. Sci. 
        Rep. {\bf 10}, 277 (1989)
\bibitem{hanbucken}M. Hanbucken, M. Futamoto, and J. A. Venables,  Surf. 
         Sci. {\bf 147}, 433 (1984)
\bibitem{loenen}E. J. van Leonen, M. Iwami, R. M. Tromp and J. F. van der 
        Veen,  Surf. Sci. {\bf 137}, 1 (1984)
\bibitem{chambers}G. P. Chambers and B. D. Sartwell,  Surf. Sci. {\bf 218}, 
        55 (1989)
\bibitem{gavioli}L. Gavioli, K. R. Kimberlin, M. C. Tringides, J. F. 
        Wendelken and Z. Zhang, Phys. Rev. Lett. {\bf 82}, 129 (1999)
\bibitem{zhang1} Z. Y. Zhang, Q. Niu and C.-K. Shih, Phys. Rev. Lett. 
        {\bf 80}, 5381 (1998)
\bibitem{suo}Z. G. Suo and Z. Y. Zhang, Phys. Rev. {\bf B58}, 5116
\bibitem{22} L. Huang, S. J. Chey and J. H. Weaver, Surf. Sci. {\bf 416}, 
         L1101 (1998)
\bibitem{su}W. B. Su, S. H. Chang, W. B. Jian, C. S. Chang, L. J. Chen 
        and T. T. Tsong, Phys. Rev. Lett. {\bf 86}, 5116 (2001)
\bibitem{okamota} H. Okamota, D. Chen and T. Yamada, Phys. Rev. Lett. 
        {\bf 89}, 256101-1(2002)
\bibitem{dipak} D. K. Goswami, B. Satpati, P. V. Satyam and B. N. Dev, 
        Curr. Sci. {\bf 84}, 903 (2003)
\bibitem{luh} D.-A. Luh, T. Miller, J. J. Paggel, M. Y. Chou and T.-C. 
        Chiang, Science {\bf 292}, 1131 (2001)
\bibitem{wei} C. M. Wei and M. Y. Chou, Phys. Rev. Lett. {\bf 68}, 125406 
         (2003); Phys. Rev. {\bf B66}, 233408 (2002)
\bibitem{altfeder}I. B. Altfeder, K. A. Matveev adn D. M. Chen, Phys. Rev. 
         Lett. {\bf 78}, 2815 (1997)
\bibitem{kmc}S. Liu, Z. Zhang, G. Comsa and H. Metiu, Phys. Rev. Lett. 
         {\bf 71}, 2967 (1993)
\bibitem{bartels}W. J. Bartels, J. Hornstra and D. J. W. Lobeek, Acta. 
         Cryst. {\bf A42}, 539 (1986)
\bibitem{luo} E. Z. Luo, S. Heun, M. Kennedy, J. Wollschlaeger and M. Henzler, 
         Phys. Rev. {\bf B49}, 4858 (1994)
\bibitem{sundar1}B. Sundaravel, A. K. Das, S. K. Ghose, K. Sekar and B. N. Dev, 
         Appl. Surf. Sci. {\bf 137}, 11 (1999) and references therein.
\bibitem{legoues}F. K. LeGoues, M. Liehr, M. Renier and W. Krakow, Philos. Mag. 
        {\bf B57}, 179 (1988)
\bibitem{zur} A. Zur and T. C. McGill, J. Appl. Phys. {\bf 55}, 378 (1984)
\bibitem{web}http://www.webelements.com
\bibitem{ino}S. Ino, T. Yamanaka and S. Ito, Surf. Sci. {\bf 283}, 319 (1983)
\bibitem{sundar2}B. Sundaravel, A. K. Das, S. K. Ghose, B. Rout and 
        B. N. Dev (unpublished)
\end{thebibliography}
 \end{document}